\documentclass[12pt]{article}
\usepackage[left=4cm, right=3cm, top=4cm]{geometry}

\usepackage{amssymb}
\usepackage{amsthm}

\usepackage[figuresright]{rotating}

\usepackage{epstopdf}
\usepackage{graphics}
\usepackage{graphicx}
\usepackage{slashed}

\begin{document}





	\begin{center}
		{\bf{On the generation of fermion condensate 
		dynamically in $SU(2)$ gauge theory}}
	\end{center}

\vskip 1mm

	\begin{center}
		{Rajdeep Basak\footnote{rajdeepbasak@gmail.com} 
		and Krishnendu Mukherjee
		\footnote{kmukherjee@physics.iiests.ac.in}}\\

{Department of Physics, Indian Institute of Engineering Science and
Technology, Shibpur, Howrah-711103 }
	\end{center}

\begin{abstract}
	Renormalized fermion condensate in $SU(2)$ gauge theory has 
	been calculated in the background of static, stable gauge field 
	configuration using field strength formalism. It is observed
        that the condensate attains a negative, minimum value at low
        energy.

\end{abstract}




\section{Introduction}

Quantum Chromodynamics (QCD) is believed to be
a correct theory for strong interaction\cite{Gross1973}.
Apart from the colour confinement,
spontaneous chiral symmetry breaking is one
of the most important aspects of strong interaction,
tnhe clear understanding of which is absent till this date.
It is believed that owing to strong interaction,
quarks of opposite chiralities are attracted
to each other to form a bound state.
The non-vanishing expectation value of the fermionic operator
$\bar{\psi}\psi$ with respect to the low energy ground state of
the theory is identified as the chiral condensate.
It is believed that chiral condensate is one of the parameters
which characterizes the vacuum structure of 
QCD at low energy\cite{Shifman1979}.
The intrinsic non-perturbative nature of the
chiral condensate indicates the difficulty
of calculating it from the first principle.

We consider an $SU(2)$ gauge theory with fermion
in the field strength formalism\cite{Halpern1979}.
Since, it is possible to obtain the information about 
the non-perturbative behaviour of a theory using 
semi-classical methods\cite{Jackiw1977}, we consider
the motion of fermion in the background of a stable
classical solution of the gauge fields\cite{Basak2019}.
The fermion condensate has been obtained after performing
the integration over the fermion fields. The renormalized condensate
assumes a minimum with respect to fluctuations 
of the mass parameter of the theory. The minimum value of 
the condensate becomes vanishingly small in the 
perturbative region of the theory. 
At low energy there is a critical scale
where the value of the condensate is zero and it remains 
negative below that scale.

\section{Evaluation of $\langle\bar{\psi}\psi\rangle$}

The Partition Function for the quarks in $SU(2)$ gauge theory 
is given by
\begin{equation}
	Z[\bar{\eta}_1, \eta_2]=\int{\cal{D}}\bar{\psi}{\cal{D}}\psi e^{iS_f}.
	\label{Z1}
\end{equation}
The action
\begin{equation}
S_f = \int d^4x  
	[\bar{\psi}(x)(i\slashed{\partial}
	+\frac{g}{2}\tau^{(a)}\slashed{A}^{(a)}(x)-m)\psi(x)
+ \bar{\eta}_1(x)\psi(x)+\bar{\psi}(x)\eta_2(x)],
\end{equation}
where $\tau^{(a)}$s ($a=1,2,3$) are the Pauli matrices.
After transforming fields in momentum space
the action becomes
\begin{eqnarray}
	S_f &=&\int \frac{d^4 p}{(2\pi)^4}
	[\bar{\psi}(p)(-\slashed{p}-m)\psi(p)
	+ \bar{\eta}_1(p)\psi(p)+\bar{\psi}(p)\eta_2(p)]\nonumber\\
	& & +\int \frac{d^4 p}{(2\pi)^4}\frac{d^4 q}{(2\pi)^4}
\bar{\psi}(q)\frac{g}{2}\tau^{(a)}.\slashed{A}^{(a)}(-p+q)
\psi(p)
\end{eqnarray}
In the axial gauge $A_3^{(a)}(x)=0$, the potentials can be solved in terms of 
fields in momentum space as\cite{Basak2019}
\begin{equation}
	A^{(a)}_0(q) = -\frac{i}{q_z}X^{(a)}_3(q),~~
	A^{(a)}_1(q) = -\frac{i}{q_z}X^{(a)}_2(q),~~
	A^{(a)}_2(q) = \frac{i}{q_z}X^{(a)}_1(q),~~
	A^{(a)}_3(q)=0.
\label{solA}
\end{equation}
Fields $X^{(a)}_j$s ($j=1,2,3$) are defined as
\begin{equation}
	X^{(a)}_1=B^a_x,~~X^{(a)}_2=B^a_y,~~X^{(a)}_3=E^a_z.
\end{equation}
Assume that quarks are confined in a background gauge field 
$\bar{X}^{(a)}_j$ which is a static solution of the equation of motion 
of the effective gauge theory\cite{Basak2019}. In the strong coupling 
limit when $\Delta\ll 1$, the background takes the form
\begin{equation}
	\bar{X}^{(a)}_{js}(q) = (2\pi)^4
\sqrt{\pi}\phi\Delta\delta^{(3)}(q_\perp)
\delta(q_z-\Delta-\zeta)\delta^{aj} e^{is\alpha_j},
	\label{Xbar}
\end{equation}
where,
\begin{equation}
s=\pm,\,\,\,\alpha_j= \frac{\pi}{6}(\delta^{j1}-\delta^{j2})
\end{equation}
and $\zeta$ is a small positive number.
We expand $S_f$ about $\bar{X}$ and drop terms to order $X-\bar{X}$.
Then we compute $S_f$ at $\bar{X}$ using eq.(\ref{Xbar}). We have 
assumed that the variation of $\psi(p)$ over the scale $\Delta$ ($\ll 1$)
is very small. Finally taking the limit $\zeta\rightarrow 0$ we obtain  
\begin{equation}
	S_f=\int\frac{d^4p}{(2\pi)^4}[\bar{\psi}(p)S^{-1}(p)\psi(p)
	+\bar{\eta}_1(p)\psi(p)+\bar{\psi}(p)\eta_2(p)+0(X(p)-\bar{X}(p))]
\end{equation}
where
\begin{eqnarray}
	S^{-1}(p) &=&-\slashed{p}-m+i\phi_1
	\tau^{(a)}(\delta^{a3}\gamma^0
	+\epsilon^{ja3}\gamma^j)\cos\theta_a,\label{Sinverse}\\
	\phi_1 &=& \sqrt{\pi}g\phi.\label{phi1}
\end{eqnarray}
Upon substitution of $S_f$ in eq.(\ref{Z1}), we integrate over the fields
and obtain the partition function 
\begin{equation}
	Z[\bar{\eta}_1, \eta_2]
	=Det[S^{-1}]\exp\left\{-i\int\frac{d^4p}{(2\pi)^4}
	\bar{\eta}_1(p) S(p)\eta_2(p)\right\} + 0(X-\bar{X}^2).
\end{equation}

The fermion condensate
\begin{eqnarray}
\langle\bar{\psi}\psi\rangle
	&=&\int\frac{d^4p}{(2\pi)^4}\frac{d^4q}{(2\pi)^4}
	\langle\bar{\psi}(p)\psi(q)\rangle\nonumber\\
	&=&\int\frac{d^4p}{(2\pi)^4}\frac{d^4q}{(2\pi)^4}
	\left(\frac{(2\pi)^4}{i}\right)^2
	\frac{\delta^2}{\delta\bar{\eta}_1(q)
	\delta\eta_2(p)} \ln Z\mid_{\bar{\eta}_1=\eta_2 =0}\nonumber\\
	&=& -i\int\frac{d^4p}{(2\pi)^4} tr S(p),
	\label{c1}
\end{eqnarray}
where $tr$ means the trace over the colour and the spinor spaces.
We write
\begin{equation}
	S^{-1}(p)=Q_0(p)+Q_1(p),
\end{equation}
where
\begin{eqnarray}
	Q_0(p) &=& -\slashed{p}-m,\\
	Q_1(p) &=& i\phi_1 \tau^{(a)}(\delta^{a3}\gamma^0
        +\epsilon^{ja3}\gamma^j)\cos\theta_a.
\end{eqnarray}
It was observed\cite{Basak2019} that $\phi=\pm b_1\Delta/\alpha_s^{a_1}$ 
in the strong coupling region, where $a_1=1.5\pm 0.003$ and
$b_1=0.00261\pm 3.0\times 10^{-6}$. It was also observed that 
$\Delta$ will remain less than $1$ in the region $1\le\alpha_s < 3069$.
So, $\phi_1=\sqrt{\pi}g\phi=\pm 1.64\times 10^{-2}\Delta/\alpha_s$ and 
it will remain small in the given region of $\alpha_s$. Consequently, 
we can treat $Q_1$ as a perturbation over $Q_0$ and obtain $S(p)$ to order
the $\phi_1^2$. Then $S(p)$ reads as,
\begin{equation}
	S(p)=Q_0^{-1}(p)-Q_0^{-1}(p)Q_1(p)Q_0^{-1}(p)
	+Q_0^{-1}(p)Q_1(p)Q_0^{-1}(p)Q_1(p)Q_0^{-1}(p)+0(\phi_1^3).
\end{equation}
The integral in eq.(\ref{c1}) is divergent in the ultraviolet limit.
We regularize the integral by going over to $d$ dimension, where 
$d=4-\epsilon$ ($\epsilon > 0$). We evaluate the trace of $S(p)$ 
in $d$ dimension and obtain the condensate upon substitution of the 
result in eq.(\ref{c1}) as
\begin{eqnarray}
	\langle\bar{\psi}\psi\rangle
	&=& -i\left(\frac{\mu}{2\pi}\right)^\epsilon(4-\epsilon)m
	\int\frac{d^dp}{(2\pi)^d}\Big[\frac{2}{p^2-m^2}
	+\frac{\phi_1^2}{(p^2-m^2)^3}\{-9(1-\epsilon)p^2\nonumber\\
	& &-20p_0^2 +(7-3\epsilon)m^2\}\Big] + 0(\phi_1^4),
\end{eqnarray}
where $\mu$ is a constant of mass dimension $1$. We evaluate 
the integrals\cite{Collins} and obtain
\begin{eqnarray}
	\langle\bar{\psi}\psi\rangle
	&=& \frac{\pi^2m}{2\epsilon}(32m^2-49\phi_1^2)
	+4\pi^2(1-2\gamma_E)m^3
	+\frac{\pi^2}{8}(17+98\gamma_E)\phi_1^2m\nonumber\\
	& &+\frac{\pi^2}{4}(49\phi_1^2-32m^2)
	m\ln\left(\frac{\pi m^2}{\mu^2}\right)
	+0(\epsilon, \phi_1^4).
\end{eqnarray}
The first term is divergent in the $\epsilon\rightarrow 0$. 
We cancel this divergence by adding counterterms to the 
action. Then, we obtain the renormalized value in the minimal 
subtraction scheme as
\begin{equation}
	\langle\bar{\psi}\psi\rangle_R
	=\sqrt{\pi}\mu^3 F(x),
\end{equation}
where
\begin{eqnarray}
	F(x) &=& 4(1-2\gamma_E)x^3
	+\frac{\lambda^2}{8}(17+98\gamma_E)x
	+\frac{1}{2}(49\lambda^2-32x^2)x\ln{x},\\
	x &=& \frac{\sqrt{\pi} m}{\mu},
	\,\,\,\,\,\,\,\,\,\,\,
	\lambda = \frac{\sqrt{\pi} \phi_1}{\mu}.
\end{eqnarray}
\begin{figure}[!th]
\begin{center}
\includegraphics[scale=0.4]{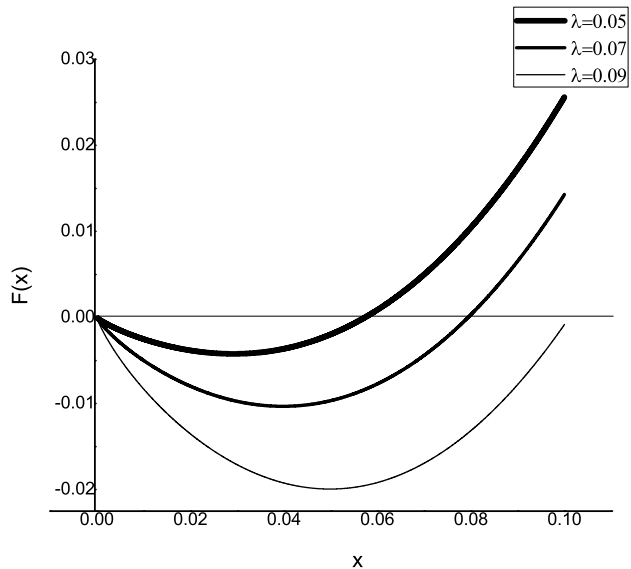}
\end{center}
\vskip -0.45in	
\caption[]{(Color online) Plot of $F(x)$ versus
$x$ for different values of $\lambda$.} 
\label{figFvsx}
\end{figure}
$F(x)$ is plotted with respect to $x$ for different values of 
$\lambda$. $F(x)$ has a minimum for $\lambda>0$. 
It indicates that the value of the condensate  
remains stable under small fluctuation of the parameter $x$ about 
its minimum value. If the minimum value of $F(x)$ occurs at $x=x_0$, then 
Fig.\ref{figFvsx} suggests that $x_0$ increases with $\lambda$ 
and $F(x_0) < 0$ for $\lambda>0$. Our numerical evaluation 
suggests that $x_0=a\lambda+b$, where $a=0.59\pm 0.001$ 
and $b=4.6\times 10^{-4}\pm 2.4\times 10^{-5}$.
The renormalized, stable value of the condensate reads as
\begin{equation}
	\langle\bar{\psi}\psi\rangle_0
	=\sqrt{\pi}\mu^3 F(x_0),
	\label{c2}
\end{equation}
At scale $\mu$, the parameter 
\begin{equation}
	\lambda(\mu)=\pm 2.9\times 10^{-2}\frac{\Delta}{\mu\alpha_s(\mu)}.
\end{equation}
From perturbative evaluation we know that $$\mu\alpha_s(\mu)
=\mu\alpha_s(\mu_0)
(1+\frac{c}{2\pi}\alpha_s(\mu_0)\ln(\mu/\mu_0))^{-1},$$ where
$c=20/3$ for $SU(2)$ gauge theory. As $\mu$ increases 
$\mu\alpha_s(\mu)$ increases 
and consequently $\lambda(\mu)$ decreases with the increase of $\mu$.
So, as $\mu$ decreases, $\lambda$ increases and the non-perturbative
region of the theory can be probed for relatively larger values of 
$\lambda$. Since $F(x_0)\rightarrow 0$ as $\lambda\rightarrow 0$,
$\langle\bar{\psi}\psi\rangle_0$ becomes vanishingly 
small at very high energy.
It indicates that $\langle\bar{\psi}\psi\rangle_0$ versus $\mu$
curve will have a minimum below the $\mu$ axis at low value
of $\mu$ and it goes to zero as $\mu$ tends to infinity.

\section{Conclusion}

We have used the field strength formalism for $SU(2)$ gauge theory 
to compute the fermion condensate in the stable background of the gauge
fields in terms of a mass parameter of the theory. 
The condensate attains its minimum at a particular value 
of the parameter and it remains stable under small fluctuation 
of the parameter around that minimum value.
The value of the condensate at the minimum  is taken as the stable 
value $\langle\bar{\psi}\psi\rangle_0$. It has been observed that this
value becomes vanishingly small at very high energy and it goes 
to zero at very low energy. 
It has been observed that
$\langle\bar{\psi}\psi\rangle_0\rightarrow 0^-$ at very high 
and very low energies. Therefore, $\langle\bar{\psi}\psi\rangle_0$
attains a negative, minimum value at relatively low energy.


\begin{thebibliography}{00}

\bibitem{Gross1973}D. J. Gross, F. Wilczek, Phys. Rev. D 8
(1973) 10.
\bibitem{Shifman1979}M. A. Shifman, A. I. Vainshtein, V. I. Zhakharov,
        Nucl. Phys. B 147 (1979) 385; (1979) 448.
\bibitem{Halpern1979}M. B. Halpern, Phys. Rev. D 19 (1979) 517.
\bibitem{Jackiw1977}R. Jackiw, Rev. Mod. Phys. 49 (1977) 681.
\bibitem{Basak2019}R. Basak and K. Mukherjee
Phys. Lett. B 789 (2019) 341.
\bibitem{Collins}John C. Collins {\it{Renormalization}}, Cambridge
Monographs on Mathematical Physics(Cambridge University Press,
Cambridge, 1984).


\end{thebibliography}
\end{document}